# Evolution of genomes in the hybridogenetic populations modelled by the Penna model

Mateusz Kula[1], Katarzyna Bońkowska[1], Maria Ogielska[2], Piotr Kierzkowski[2], Anna Zaleśna[2], Stanisław Cebrat[1]

[1] Department of Genomics, Faculty of Biotechnology, University of Wrocław, ul. Przybyszewskiego 63/77, 51-148 Wrocław, Poland

[2] Department of Biology and Conservation of Vertebrates, Zoological Institute, Faculty of Biological Sciences, University of Wrocław, ul. Sienkiewicza 21, 50-335 Wrocław, Poland.

e-mail addresses:

Mateusz Kula – toczka@smorfland.uni.wroc.pl

Katarzyna Bońkowska - kasia@smorfland.uni.wroc.pl

Maria Ogielska - ogielska@biol.uni.wroc.pl

Piotr Kierzkowski - piotrk@biol.uni.wroc.pl

Anna Zaleśna - azal@biol.uni.wroc.pl

Stanisław Cebrat – cebrat@smorfland.uni.wroc.pl

*Correspondence to:*

S. Cebrat; *e-mail:* cebrat@smorfland.uni.wroc.pl;

tel.:+48-71-3756-303;

fax:+48-71-3252-151

*Web site:* http://smORFland.uni.wroc.pl

## Abstract

**Background:**

Hybridogenesis is a very interesting example of reproduction which seems to integrate the sexual and clonal processes in one system. In a case of frogs, described in the paper, two parental species – *Rana lessonae* and *Rana ridibunda* can form fertile hybrid individuals – *Rana esculenta*. Hybrid individuals eliminate one parental haplotype from their germ line cells before meiosis (end before recombination) which implicates clonal reproduction of the haplotype transferred to the gametes. All three "species" are called "complex species". To study the evolution of genomes in the hybridogenetic fraction of this complex species we have used the Monte Carlo based model rendering the age structured populations. The model enables the analysis of distribution of defective alleles in the individual genomes as well as in the genetic pool of the whole populations.

**Results:**

We have shown that longer isolation of hybrids' populations leads to the speciation through emerging the specific sets of complementing haplotypes in their genetic pool. The fraction of defective alleles increases but the defects are complemented in the heterozygous loci. Nevertheless, even small supply of new hybrids generated by the two parental species or crossbreeding between hybrids and one of the parental species prevents the speciation and changes the strategy of the genome evolution from the complementing to the purifying Darwinian selection.

**Conclusions:**

The computer modeling of hybridogenesis has shown that the hybridogenetic populations could be stable and they are not under the pressure of the Muller's ratchet effect. Since the interplay between the intragenomic recombination rate and the inbreeding coefficient is very important in the process of the genome evolution, the reproduction system exploiting the hybridogenesis can be considered as self-control of the recombination frequency in the natural populations.

**Background**

**Hybridization and alternative modes of sexual reproduction**

Hybridization may in some cases lead to speciation, see [1,2,3] for review . Viable interspecies hybrids are heterozygous in more alleles than the parental species and thereby may display the effect of heterosis. In such a case heterozygosity leads to expression of a phenotype that better tolerates environmental stresses and successfully competes with the related species. In case when a hybrid is fertile, the effect of heterosis will gradually disappear in the next generations owing to recombination. However, in most cases the hybrids are sterile or have low fertility caused by disturbances during meiosis. The only way to escape from such disturbances and to maintain the new hybrid phenotype is stabilization of the genotype of F1 generation. This situation is possible only in cases when reproduction is asexual or when some unusual alternative mechanisms appear before or during gametogenesis, such as parthenogenesis, gynogenesis or hybridogenesis [4,5]. Hybridogenesis in water frogs of the *Rana esculenta* complex occurs in both sexes, but production of gametes (both sperm and ova) is clonal and recombination is very low (reviewed by Schmeller [6]). Before meiosis one of the parental genome is eliminated and thereby gametes produced by a hybrid are clonal; when fertilized by gametes of a related species, the progeny is semiclonal.

Hybridogenesis is only one aspect of the highly complicated reproduction in *Poeciliopsis* fishes, including poplipolidization and gynogenesis, see [7,8] for review. The other fish species with hybridogenetic reproduction belong to the *Leuciscus alburnoides* complex [9] and *Phoxinus eos-neogaeus* complex [10]. Hybridogenesis was also reported in invertebrate stick insects of the *Bacillus rosius-grandii-benazzii* complex [11,12,13].

*Rana esculenta* complex (subgenus *Pelophylax*) is common and widely distributed in the Palearctic. According to the model of hybridogenesis, *R. lessonae* (genotype LL), *R. ridibunda* (genotype RR) are species, whereas *R. esculenta* (genotype RL) is a hybrid with *lessonae* (L) and *ridibunda* (R) haplotypes (for review see [14]). Direct hybridization between *R. ridibunda* and *R. lessonae* in the nature is rare because the ecological preferences of the two parental species are different [15,16,17,18,19]. For this reason they rarely coexist in the same territory. Hybridization is possible in three basic kinds of mixed populations, which in the most cases they form genetic systems with predictable pattern of reproduction: *lessonae-esculenta* (L-E), *ridibunda-esculenta* (R-E), and homotypic *esculenta-esculenta* (E-E) [20,21,22,23].

**Modeling the hybridogenesis in the *Rana esculenta* system**

Accumulation of mutations in hybridogenesis was recently modeled by Som et al., [24]. Authors noticed that since hybridogenesis leads to the cloning of one haplotype it has to end with genetic meltdown of the cloned haplotype considered as the Muller ratchet effect [25]. They analyzed evolution of hybridogenetic populations of different sizes and under different male to female mutation ratio. They found that under some conditions the genetic load of the hybrids' population could be even lower than of sexual parental species. Nevertheless, they used the Redfield [26] model of evolution. There are some assumptions in the model which can lead to the wrong results or wrong conclusions. The most important is a co-dominance; all defective alleles have the same effect for fitness of the individual, independently of there location i.e. two defects in the same locus (homozygous) or in different loci (heterozygous). Populations are not age structured and the genetic state of population is estimated in the probability distribution terms rather instead of distribution of genetic status of individuals. Additionally, in this kind of modeling, recombination is introduced in such a way, that mutated alleles from heterozygous loci are introduced into the gamete with probability 0.5. This assumption of the model connects directly the probability of recombination between the haplotypes with the genetic load. To avoid the effects of these assumption we have used the Monte Carlo based Penna model where individuals were represented by their diploid genomes with two bit strings representing haplotypes [27].

We have modeled a specific strategy of hybridogenetic reproduction exploited by two species *R. lessonae* and *R. ridibunda*) which can crossbreed and form the surviving and fertile hybrids (*R. esculenta*). Hybrids can crossbreed within the system with other hybrids or with individuals from each of the parental species but there is no recombination between parental haplotypes during gametogenesis. In the gonads of hybrid organisms one parental haplotype is lost and they produce gametes with the other haplotype only. To model this type of reproduction we have used the Penna model [27]. It supports the Medawar's hypothesis [28] assuming that in the genetic pool of populations the defective genes expressed late during the lifespan of organisms are accumulated which renders the ageing processes. The defective genes expressed before the minimum reproduction age are eliminated by the Darwinian purifying selection. This model has been used successfully for modeling the evolution of the age structured populations (see Stauffer et al., [29] for review). It quantitatively describes a lot of phenomena like trade off between the fecundity and the lifespan [30], differences in the mortality of men and women [31], menopause [32], the effect of random death on genetic pool of populations [33], higher mortality of the youngest individuals [34], and many other biological processes connected with the population evolution [35-39]. The most interesting feature of the model, from the point of view of the hybridogenesis studies, is the possibility of analyzing the changes of genetic structure of the single individuals as well as of the whole genetic pool of population during the evolution. The model predicts also the interplay between the inbreeding and the intragenomic recombination rate during the population evolution and sympatric speciation [40-43]. In the conditions of high inbreeding – when the panmictic populations are very small or in the larger populations of genetically highly related individuals, the probability of meeting two identical chromosomes or their parts in one zygote is relatively high and phenotypic expression of defective homozygous loci is also high. In such conditions there are two possible strategy of reproduction. One of them is the purifying selection – demanding relatively high intragenomic recombination rate. The other strategy – under low recombination rate is based on completing the genomes of the offspring of two complementing haplotypes. Such genomes can survive under much higher genetic load [42, 43].

It seems that water frogs live under different and fluctuating inbreeding conditions and that is why they exploit at least temporarily the hybridogenetic mode of reproduction which could be favorable under such conditions. To check if hybridogenesis could be evolutionary profitable, under what conditions and if the hybrids obligatory end up in the genetic melt down, we have used a Monte Carlo based model.

**Model**

**Standard version**. We have used diploid version of the Penna model. Each individual is represented by its genome composed of two haplotypes – bitstrings each 64 bits long. Each position in the bitstring (locus) is occupied by a bit set to 0 or to 1 corresponding to the wild or defective version of an allele, respectively. We have assumed that all defective alleles are recessive; the defective phenotype is expressed only if both alleles at the corresponding locus in both bitstrings are defective. The characteristic feature of the model is a chronological switching on the genes. Organisms die when a declared number of phenotypic defects are switched on (threshold T). If organism survives until the minimum reproduction age (R) it can reproduce. Before the reproduction, a female replicates its haplotypes and introduces with probability M a new mutation into a copy of each haplotype at the randomly chosen locus. Mutation changes the bit set to 0 into bit set to 1, if the value of the bit has been already 1 it stays 1 – there are no reversions. The copies of the two parental haplotypes recombine at the randomly chosen point exchanging their arms in the process mimicking the meiotic crossover. Each of the resulting haplotypes represents a gamete. The female looks for a male individual at the reproduction age, which produces its gamete in the same way. Male and female gametes form a zygote and its sex is set with equal probability to male or female. During the one iteration (Monte Carlo step – MCs) each individual is checked for survival after switching on the alleles in the consecutive locus, eventually dies if it reaches threshold T. The survivor reproduces if it is at the reproduction age. In the model, individuals can die because of their genetic status, randomly or after reaching the maximum age which equals the length of the bitstrings. In fact they never reach the maximum age dying earlier because of the genetic death. Random death is set for newborns only by the logistic equation of Verhulst factor : $V = 1 - N_t/N_{max}$; where V corresponds to the probability of the offspring survival, $N_t$ is the current size of population and $N_{max}$ is the capacity of the environment. This function keeps the size of population below the maximum capacity of the environment. When the Verhulst factor eliminates only newborns, the older individuals can be eliminated only by the genetic death. It is also possible to introduce the random death into the older fractions of populations, then the evolutionary costs of elimination the defective genes are higher and in such cases the genetic status of population is worse [33].

Thus, there are a few free parameters of the model, with their values assumed in the simulations in the brackets:

L – length of the bit strings – number of virtual genes in one haplotype [64],

M – mutation rate, number or probability of new mutations introduced during the haplotype replication [1 per haplotype per replication],

C – recombination rate, probability of crossing over between haplotypes during the gamete production [1 in case of parental species, 0 for hybrids],

B – birth rate, the number of the offspring produced by one female at the reproduction age during one MCs [1 if not stated otherwise in the text],

R – the minimum reproduction age [40 bits expressed in two MCs],

T – the number of defective phenotypic traits which kill the individual [1].

In the case of modeling the hybridogenesis, the time of evolution measured in the number of Monte Carlo steps or generations is a very important additional parameter affecting the final status of populations.

**Models of hybridogenesis**.

In all versions described below we have assumed that populations of parental species and hybrids do not compete. Such conditions are introduced by declaring that the Verhulst factor operates separately for each coevolving population.

1. In the first version of hybridogenesis we have used a simple modification of the standard Penna model. The individuals of the hybrids' population can crossbreed only inside this population without any contact with parental species. The significant difference between this version and the standard model is in switching off the recombination between haplotypes during the gametogenesis. The initial population of hybrids is generated by crossbreeding of two independently evolved parental species being already in the equilibrium.

2. In the second version, hybrids can crossbreed within their own population, like in the first version, but into the population of hybrids the "new" hybrids are introduced with a declared probability. These new hybrids are produced by hybridization the two parental species coevolving in the noncompetitive conditions. Thus, in the whole hybrid population we have "old" hybrids and the "newcomers" which can freely interbreed.

3. Hybrids can crossbreed with only one parental species with a declared probability. We have assumed that only hybrid females can look for males of parental species to reproduce.

4. Like in version 3, but hybrids can crossbreed with both parental species with declared probabilities.

In the versions 2, 3 and 4 the initial population of hybrids is formed like in the version 1 and the results of simulations are shown starting from the very beginning of the hybrid population evolution. The data for parental species are presented at their equilibrium.

**Results**

In these simulations we have analyzed the effect of different strategies of reproduction on the genetic pool of populations. When the Verhulst factor controls the population size and the birthrate is relatively high, the population size itself is a poor discriminatory trait of the quality of population. Much better quantitative features are: the fractions of defective genes and their distribution (Fig. 1), the age distribution of population (Fig. 2) and/or the mortality described as the logarithm of fraction of population dying in a given period of life plotted versus the age (Gompertz plot [44]).

In the first version of simulations there is no statistically significant difference between the two parental species what is obvious, because all parameters of simulations, including the capacity of the environment (inbreeding) were identical. The hybrid population, under the same conditions was smaller than each of the parental populations (Fig. 2). But the most pronounced difference between the population of hybrids and the populations of parental species was in the genetic pool structure (Fig. 1). The fraction of defective genes in the pool of genes expressed before the minimum reproduction age in the parental populations was of the order of 0.15 while in the hybrid population it reaches 0.5. The number of genes expressed before the minimum reproduction age was set to 40. If the defective genes in this set were distributed randomly, the probability of surviving the hybrid offspring till the reproduction age would be of the order of $0.75^{40}$ which roughly equals 1 per $10^5$. In such conditions the population has to be extinct. To survive, the best strategy of reproduction is to form zygotes of two complementing haplotypes. That is why during the evolution two complementing groups of haplotypes emerge and at the equilibrium state only two haplotypes exist in the population. Moreover, this strategy of reproduction seems to be stable and the probability of the population extinction depends only on the initial size of population (fluctuations) and it is not under the pressure of the Muller ratchet any more. In Fig. 3 we have shown the evolution of Hamming distance between two haplotypes in the hybrids' genomes. In this case the Hamming distance corresponds to the number of heterozygous loci in the first 40 bits in the bitstring – genes located in these loci are expressed before the minimum reproduction age. After 10,000 MCs the distance reaches 40, which means that both haplotypes are fully complementary. This state is stable and does not change during the next 100,000 MCs (tested but not shown in this scale). Probability of the offspring survival until the minimum reproduction age depends also on the mutation rate and under the simulation parameters it is about 0.25. Thus, the population would be extinct if the total fecundity would not compensate this mortality.

**Influence of the new hybrid generation on the equilibrium state of population**

In the simulations described above the hybrid populations, once generated, evolved without any contact with the parental populations and their genetic pools. In the version described below we have assumed that the two parental species can generate new hybrids which enrich the population of earlier generated hybrids (new hybrids can be generated in the same environment or can immigrate from the remote locations). We have performed simulations with different fractions of the newly generated hybrids (called here "newcomers"). The important assumption was that there are no differences in the sexual behavior of the members of the old population and the newcomers. The separated population of hybrids has a tendency to complement the haplotypes rather than to purify them of the defective genes. In the new version, the genetic pool of population is enriched in the haplotypes which have been under purifying selection in the parental species. Results are shown in Fig. 4. The effect of the supply of new hybrids on the genetic structure of population is nonlinear. At very low fraction of newcomers hybrid population still exploits the strategy of haplotype complementation, and fraction of defective alleles is close to 0.5. When the fraction of newcomers passes 0.04 the strategy switches to the purifying selection. It is also shown in Fig. 5. where the genetic structure of hybrids is presented. Even small influx of new hybrids affects the defective alleles' distribution in the whole genetic pool.

**Backcrosses of hybrids with parental species.**

There are two different situations when the effects of backcrosses could be analyzed. In the first one, both populations (hybrids and a parental one) cannot crossbreed until they reach the equilibrium. In the second one, hybrids can be backcrossed immediately after their formation from the parental species. In the first situations the haplotypes in the hybrid's genetic pool have already formed two separate complementing sets and the probability of forming the viable organisms with parental haplotypes is extremely low. In fact they already represent different species. This scenario corresponds to suggestion of Vrijenhoek [8,46] that hybridogenesis could be considered as the first step of speciation or could lead to speciation. In the second scenario all haplotypes in the crossbreeding populations have been under purifying selection and the switching to the complementary strategy could depend on the intensity of hybrid backcrossing when comparing with the interbreeding of hybrids within their population. If the intensity is high enough, the hybrids do not switch into the complementary strategy and they keep the low fraction of defective genes like the parental species. The results are shown in Fig. 5. They are similar to the situation where new hybrids where introduced into the hybrid population.

**Discussion**

Hybridogenesis is a specific way of sexual reproduction of the complex species. Let's consider it as a method of self-control of recombination frequency between parental haplotypes. Our previous studies have shown, that interplay between the inbreeding and intragenomic recombination rate could be very important for the strategy of the genome evolution [42,43]. Under high inbreeding and low crossover rate the more effective strategy for surviving is to complement defects in haplotypes instead of elimination of defective alleles. This strategy can lead to the fast sympatric speciation which has been probably observed in the hybridogenetic complex species. In our studies we have performed simulations in the simplest possible conditions, just to show how the Monte Carlo model of age structured population can be used in such studies. In our opinion, the Penna model can be used for further studies of hybridogenetic strategy of reproduction. Previous studies suggested that one of the most important parameter in the population evolution is the size of population. In fact it is not the size itself (it can have important effect in the drift phenomena) but the inbreeding coefficient. In panmictic populations inbreeding decay very fast (exponentially) with the size of population but in case of frogs it seems to be much more complicated.

Theoretically, the reproduction potential of frog population is huge. Nevertheless, the population size is fluctuating but kept in the reasonable range. We can assume that the inbreeding fluctuates strongly in such populations then, it is advantageous to introduce some mechanism of controlling the intragenomic recombination rate – hybridogenesis could be considered as one of the possible and effective way of doing that. It could be analyzed quantitatively in the Penna model after introducing the proper parameters of reproduction. In our case the probability of giving the offspring was rather evenly distributed in the adult population and all adults form one panmictic population. Preliminary studies have shown that the possibility of crossbreeding with parents significantly changes the results.

**Conclusions**

Switching off the recombination between parental haplotypes leads to the strategy of complementing the haplotypes in the hybrid individuals. Such a strategy allows accumulation of higher fraction of defective alleles without their phenotypic expression. Nevertheless, a small fraction of new hybrids or backcrossing with parental species favours the purifying selection which keeps the fraction of defective alleles at the level of parental species.

**Authors' contributions**

SC – the idea, assumptions of the model and writing draft of the manuscript, MK and KB programming and simulations, MO, PK and AZ biological background and introduction into biology of hybridogenesis. All authors read and approved the final manuscript.

**Acknowledgements**

This work was supported by KBN grant #105/E-344/SPB. It was done in the frame of European programs: COST Action P10 and FP6 NEST – GIACS. MO was supported by KBN 3490/B/P01/2007/33

Figures
Fig. 1.

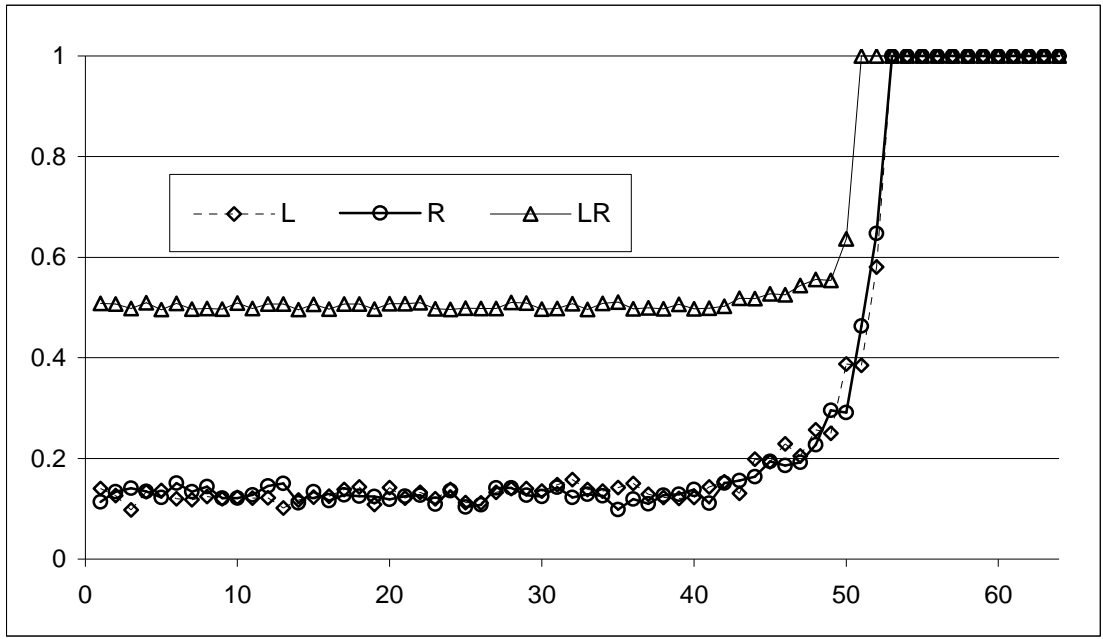

Fig. 1. Distribution of defective genes in the genomes of parental species and hybrids after 10 000 MCs. All populations evolved independently. Minimum reproduction age after switching on 40 bits in two MCs (see text for details).

Fig. 2.

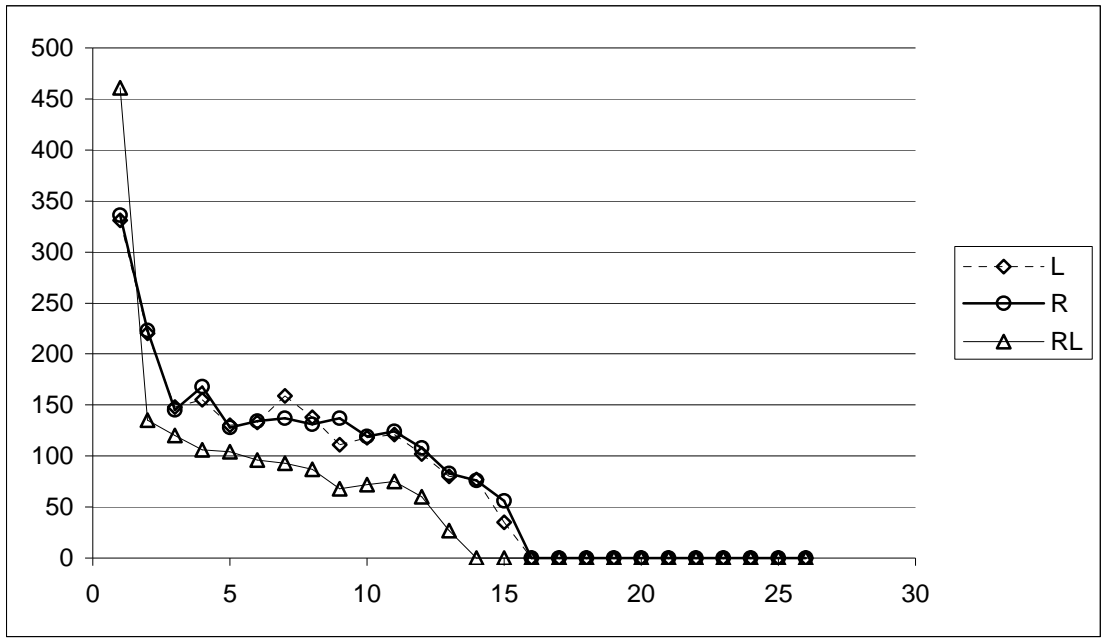

Fig. 2. Age distribution of parental lines and hybrids living in separate environments. During the first two "years" 40 bits were switched on.

Fig. 3.

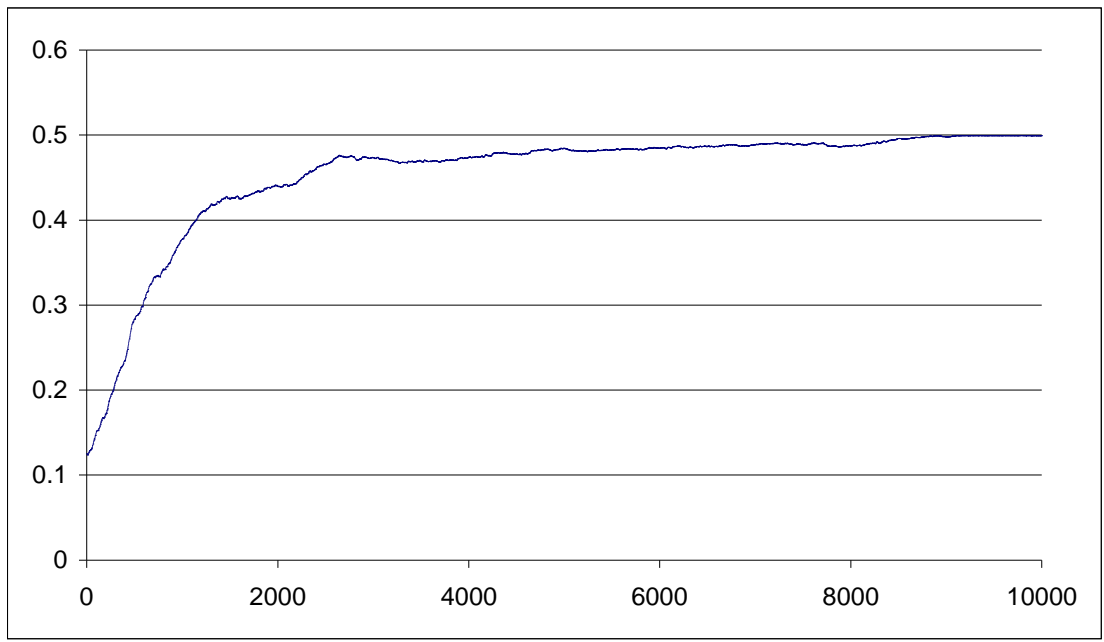

Fig. 3. Changes in the average Hamming distance between haplotypes (the first 40 bits) in the genomes of hybrids during the simulations. Hamming distance corresponds to the number of heterozygous loci in the analyzed fragment of the genome.

Fig. 4.

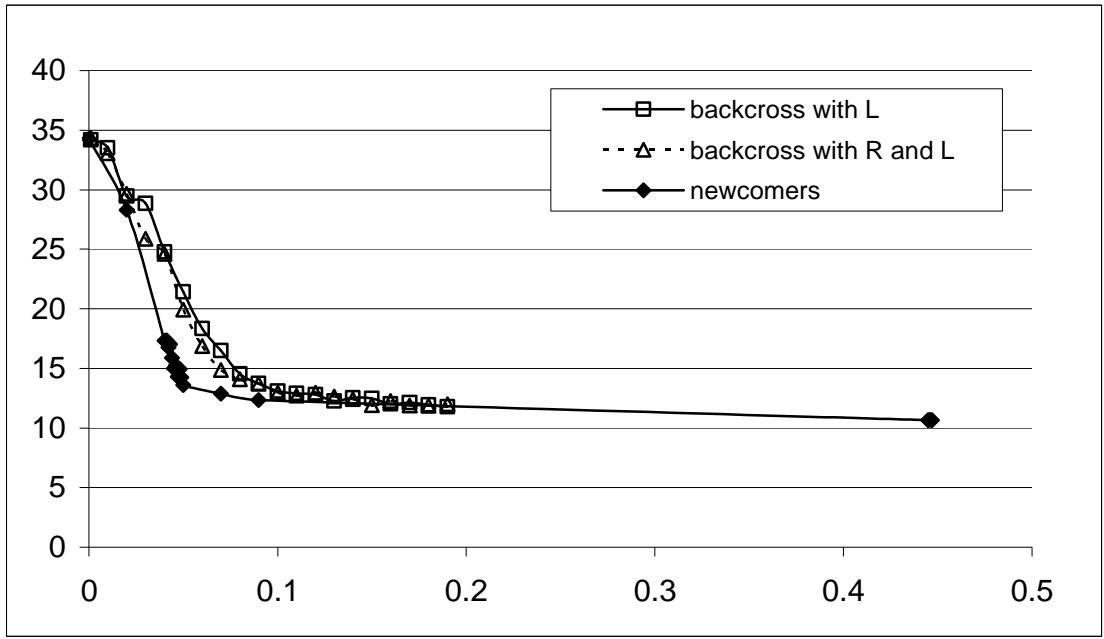

Fig. 4. The relation between the number of heterozygous loci in the first 40 bits and the fraction of new hybrids incoming to pool of hybrids (newcomers) or the frequency of backcrosses between hybrids and parental lines.

Fig. 5.

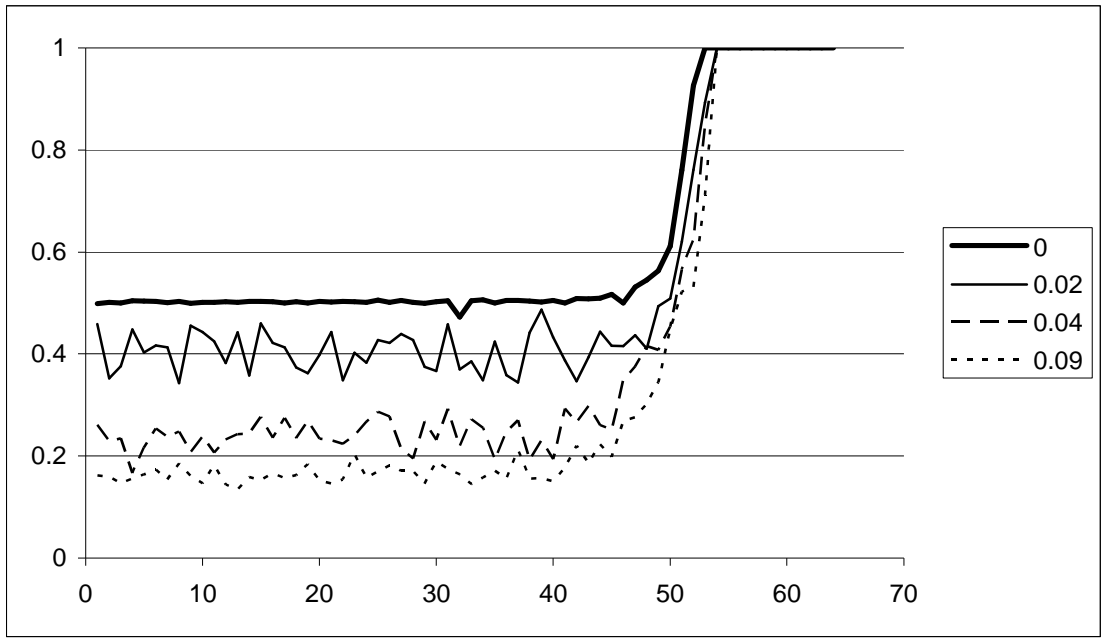

Fig. 5. Distribution of defective genes in the hybrid genomes in populations with different fraction of "newcomers".